\documentclass[aps,prl,twocolumn,preprintnumbers,10pt,showpacs,nohyper]{revtex4}
\usepackage{amsmath,bm}
\usepackage{multirow}

\usepackage[utf8]{inputenc}
\usepackage{graphicx}
\usepackage{xspace}
\usepackage{epsfig}
\usepackage{color}

\newcommand{\NNLOJET}{NNLO\protect\scalebox{0.8}{JET}\xspace}

\allowdisplaybreaks

\begin{document}

\preprint{CERN-TH-2019-054, IPPP/19/31, ZU-TH 19/19, CFTP/19-016}

\title{Triple Differential Dijet Cross Section at the LHC}

\author{A.\ Gehrmann-De Ridder$^{a,b}$, T.\ Gehrmann$^{b}$, E.W.N.\ Glover$^{c}$, A.\ Huss$^{d}$, J.\ Pires$^{e,f}$}
\affiliation{$^a$ Institute for Theoretical Physics, ETH, CH-8093 Z\"urich, Switzerland 
\\
$^b$ Department of Physics, Universit\"at Z\"urich, Winterthurerstrasse 190, CH-8057
Z\"urich, Switzerland 
\\
$^c$ Institute for Particle Physics 
Phenomenology, University of Durham, Durham DH1 3LE, United Kingdom\\
$^d$ Theoretical Physics Department,
 CERN, 1211 Geneva 23, Switzerland\\
$^e$ CFTP, Instituto Superior T\'{e}cnico, Universidade de Lisboa, P-1049-001 Lisboa, Portugal\\
$^f$ LIP, Avenida Professor Gama Pinto 2, P-1649-003 Lisboa, Portugal}


\begin{abstract}
The measurement of the triple-differential dijet production cross section as a function of the average transverse momentum $p_{T,\textrm{avg}}$, half the 
rapidity separation $y^{*}$, and the boost $y_{b}$ of the two leading jets in the event  enables a kinematical scan of the underlying parton momentum distributions. 
We compute for the first time the second-order perturbative QCD corrections to this  
triple-differential dijet cross section, at leading color in all partonic channels, thereby enabling precision studies with LHC dijet data. 
A detailed comparison with experimental CMS 8~TeV data is performed,  
demonstrating how the shape of this differential cross section probes the parton densities in different kinematical ranges. 
\end{abstract}

\maketitle

Jet production in proton--proton collisions results predominantly from hard interactions,
when two partons from the incident hadrons undergo a hard pointlike interaction and scatter at relatively large angles. As such, it is directly sensitive
to the dynamics of the pointlike strong-interaction partonic cross section and to the nonperturbative description of the internal proton structure encoded in the parton 
distribution functions (PDFs). It is the combined 
interplay between the parton--parton scattering matrix elements and the parton luminosities that determines the shape of the dijet cross section. For this reason, to fully 
exploit the wealth of available data from the LHC it is important to have a reliable and accurate theoretical prediction for the dijet cross section. More detailed information about the shape of the dijet cross section can be obtained when the latter is determined in triple-differential form~\cite{Giele:1994gf}, over a wide range in the variables that fully describe the two jet
events, namely, the transverse momentum $p_{T}$ and rapidities $y$ of the two leading jets in the event.  Triply differential measurements of dijet production have so far been performed 
by the CDF experiment~\cite{Affolder:2000ew} at the Tevatron and by the CMS experiment~\cite{Sirunyan:2017skj} at the LHC.

To use the full potential of the LHC data to achieve the ultimate experimental and theoretical precision in the way the structure of the proton is accessed, 
it is convenient to make a clever choice of variables in the dijet process, which manifest the IR safety of the observable, and that directly map the measured cross section 
to surfaces on the $(x,Q^2)$ plane where the PDFs are determined. 

In this Letter, we calculate the triple-differential dijet cross section as a function of the following three kinematical variables: 
the average transverse momentum $p_{T,\textrm{avg}}=(p_{T,1}+p_{T,2})/2$ of the two leading jets, half of their rapidity separation $y^{*}=|y_{1}-y_{2}|/2$, and the boost of the dijet system
$y_{b}=|y_{1}+y_{2}|/2$.
Our results are presented below as six distributions and compared with CMS 8~TeV data~\cite{Sirunyan:2017skj}.

Using the kinematical variables defining the triple differential dijet cross section, the longitudinal momentum fractions of the incoming partons
can, for the Born process of back-to-back jets, be written using momentum conservation,
\begin{eqnarray}
x_{1}&=&\frac{p_{T}}{\sqrt{s}}~(e^{+y_{1}}+e^{+y_{2}})=\frac{2\,p_{T,\textrm{avg}}}{\sqrt{s}}~e^{\pm y_{b}}\cosh(y^*),\nonumber\\
x_{2}&=&\frac{p_{T}}{\sqrt{s}}~(e^{-y_{1}}+e^{-y_{2}})=\frac{2\,p_{T,\textrm{avg}}}{\sqrt{s}}~e^{\mp y_{b}}\cosh(y^*).\label{eq:x12}
\end{eqnarray}
From Eq.~\eqref{eq:x12} it is clear that for small values of $y_{b}$ the dijet data probe
the configuration $x_{1}\approx x_{2}$, with the $x$-values determined by the $p_{T,\textrm{avg}}$ of the jets. Moreover, it can be seen that a variation in $y^*$ at different
$p_{T,\textrm{avg}}$, probe approximately the same $x$ at different $Q^2$.

On the other hand, a variation in the dijet cross section in $y_{b}$ is an additional handle that allows us to depart from the $x_{1}\approx x_{2}$ region and probe the scattering
of a high-$x$ parton off a low-$x$ parton. In this respect, it is noteworthy that the parton--parton subprocess matrix elements are independent of $y_{b}$. As result, for a fixed $y^{*}$ slice,
the only variation of the cross section comes from the variation of the parton densities as $y_{b}$ varies over the allowed kinematic range. The triple-differential dijet observable is therefore ideal to study
PDFs, because the choice of variables in the dijet system directly maps to surfaces on the $(x,Q^2)$ plane where the PDFs are determined, in event topologies that
are sensitive to the PDFs, while being insensitive to the matrix elements. 
\begin{figure*}[t]
  \centering
    \includegraphics[width=0.28\textwidth]{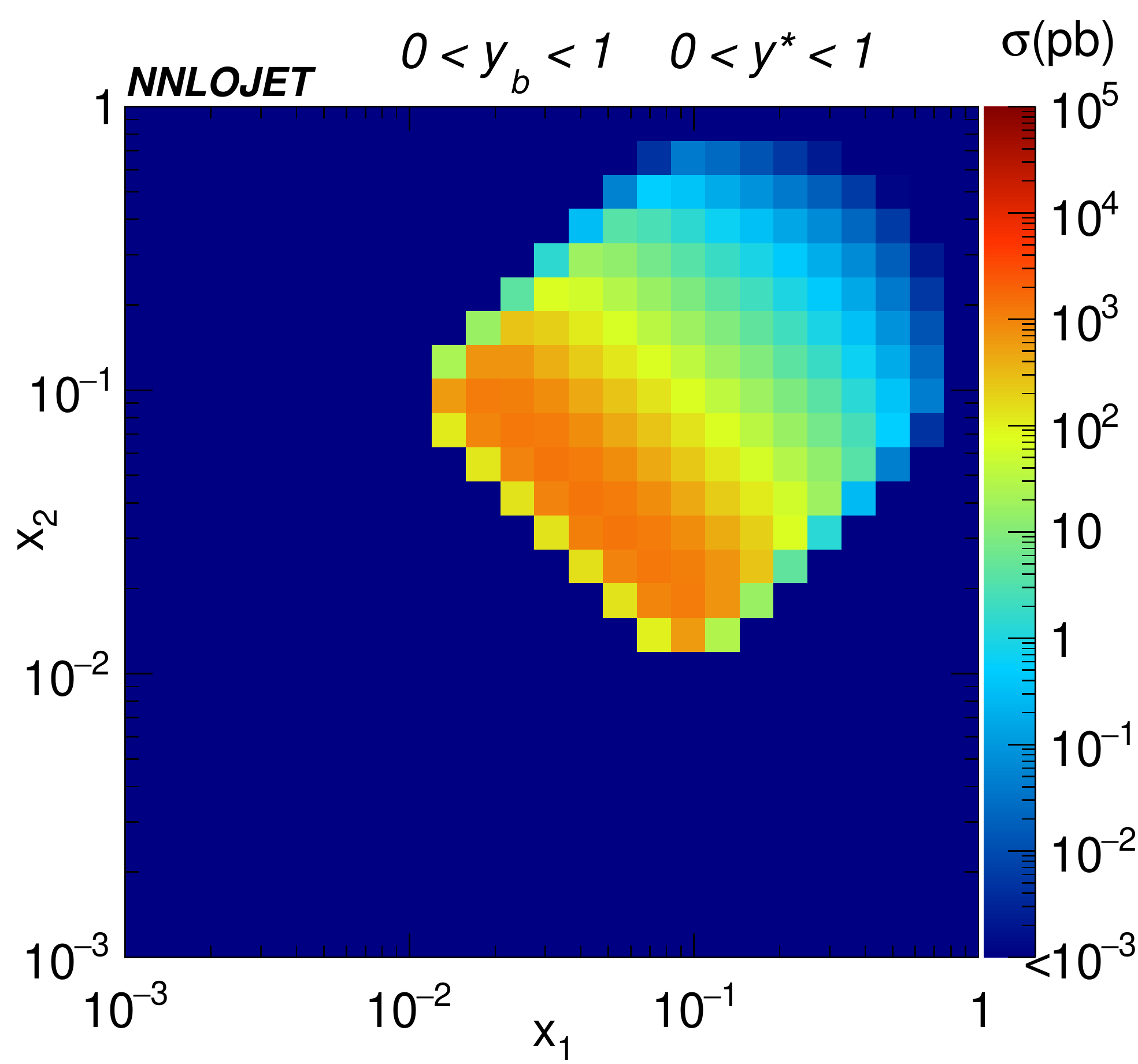}\hspace{0.5cm}
    \includegraphics[width=0.28\textwidth]{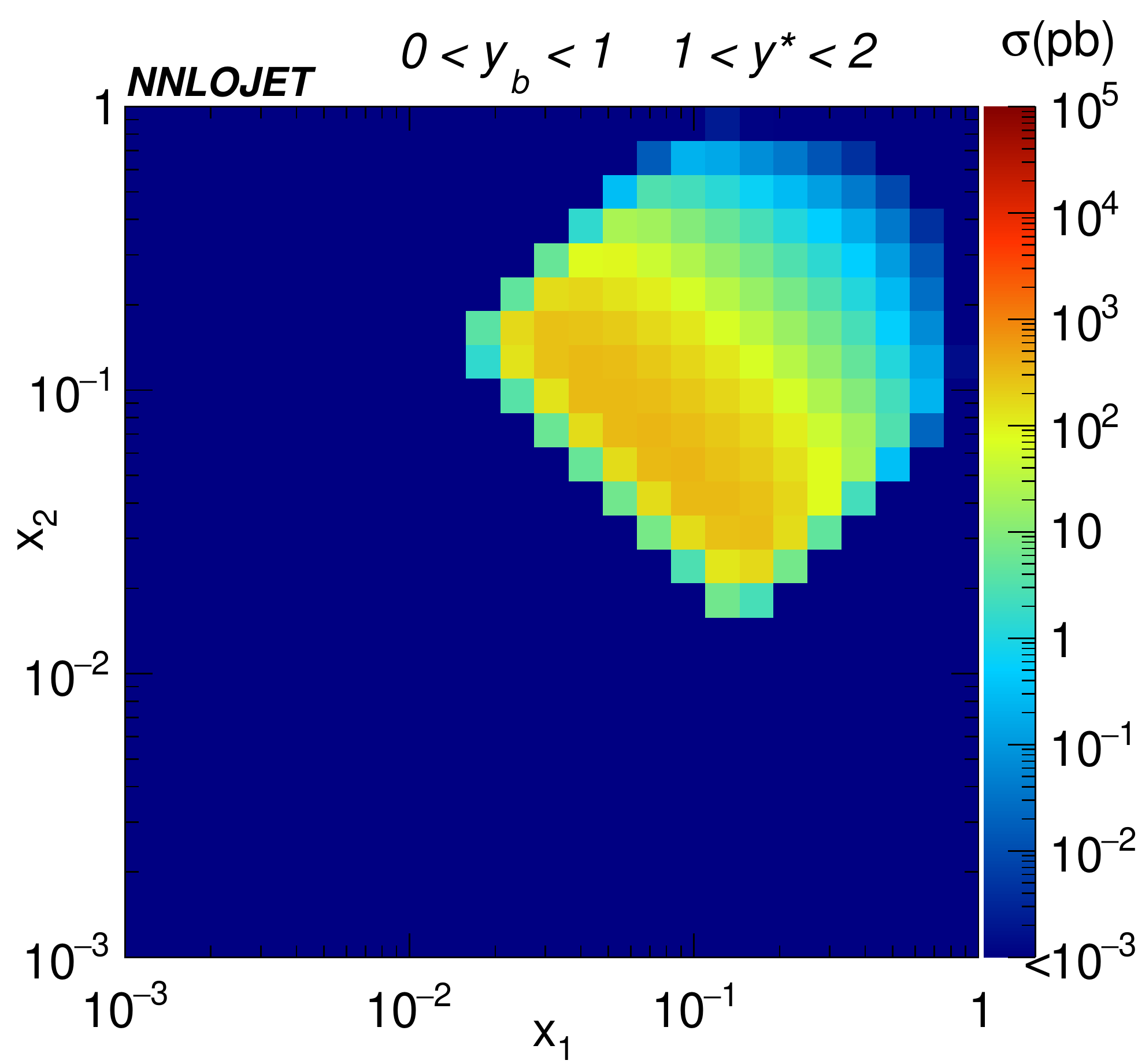}\hspace{0.5cm}
    \includegraphics[width=0.28\textwidth]{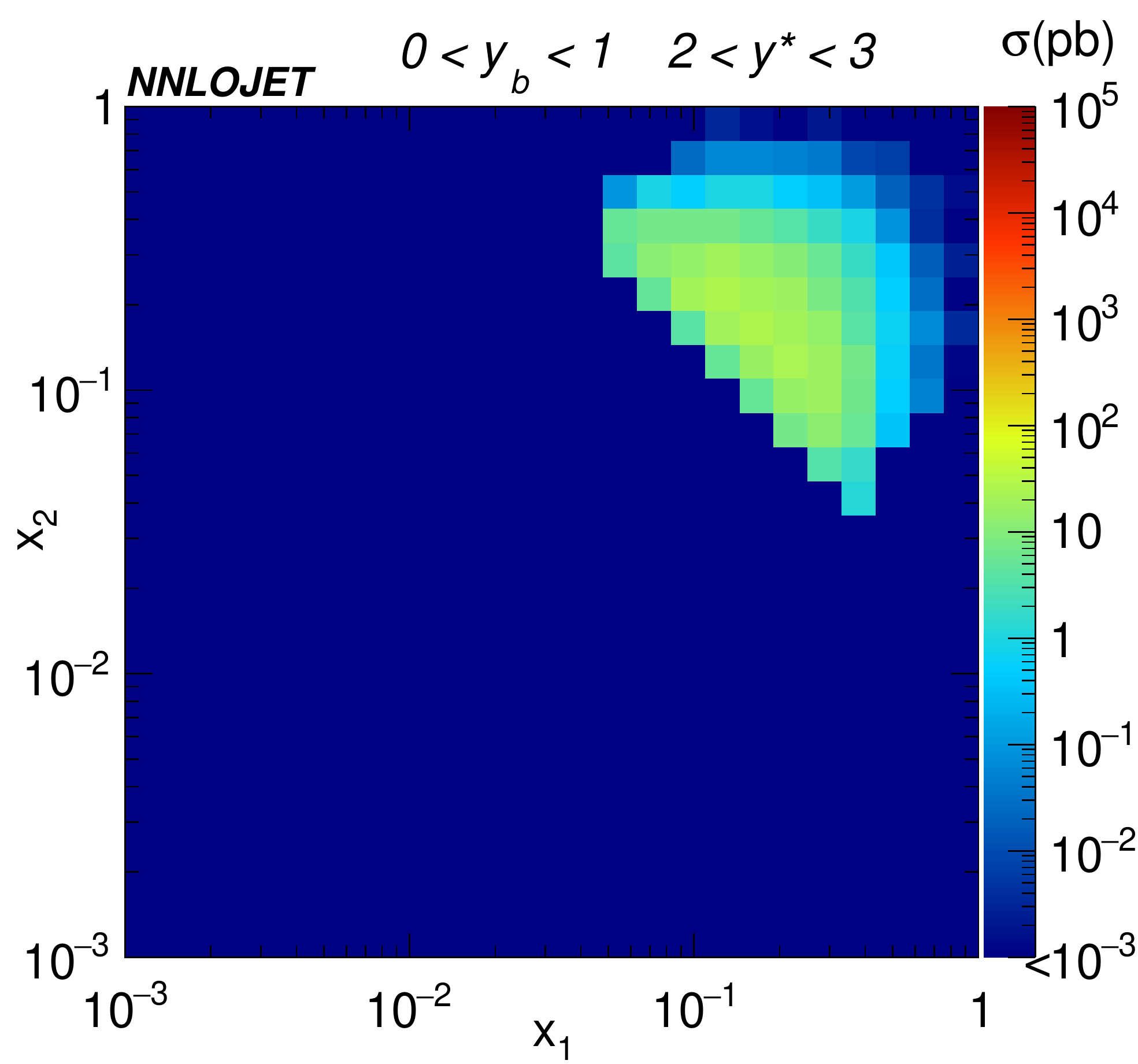}\\
    \includegraphics[width=0.28\textwidth]{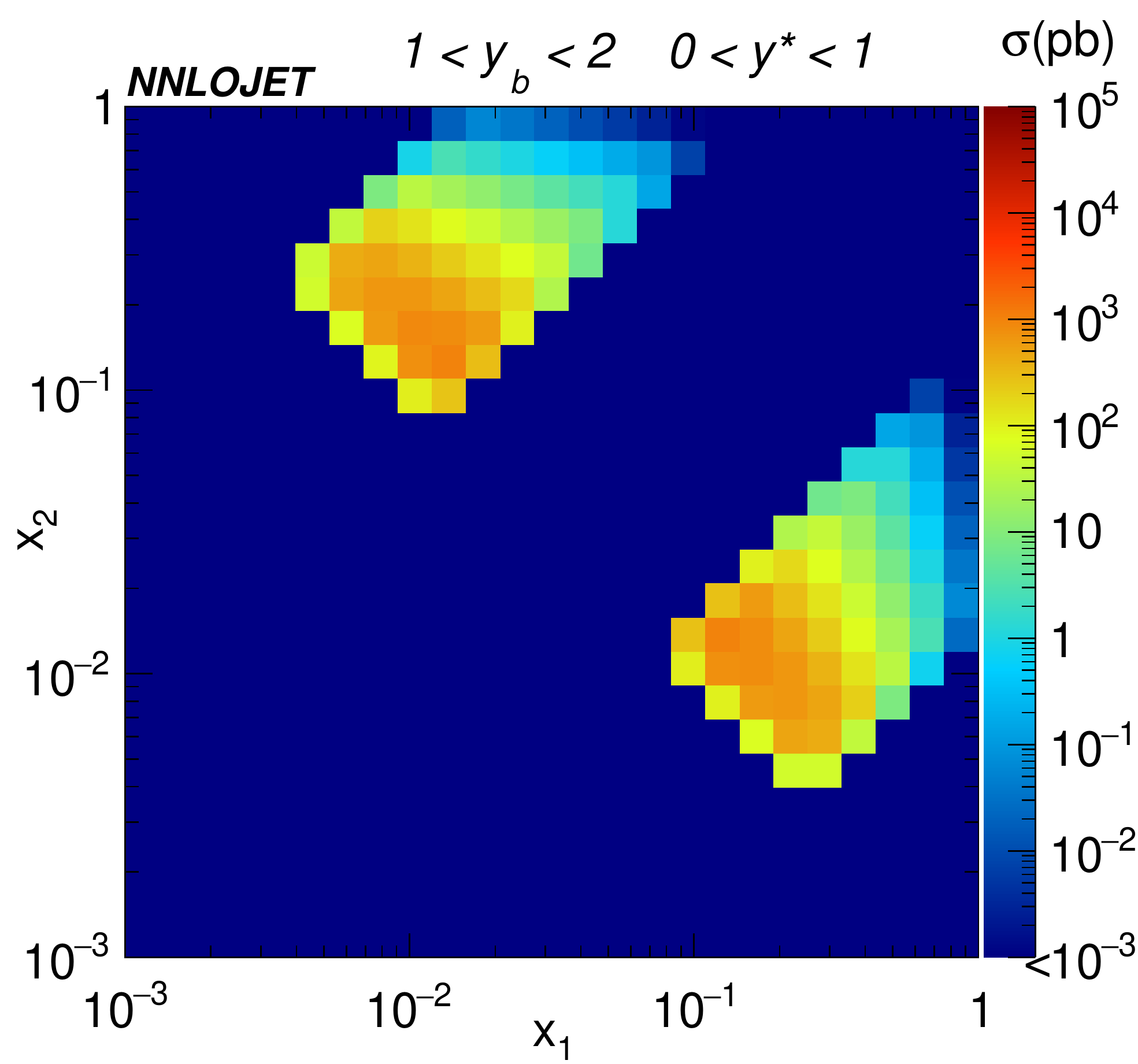}\hspace{0.5cm}
    \includegraphics[width=0.28\textwidth]{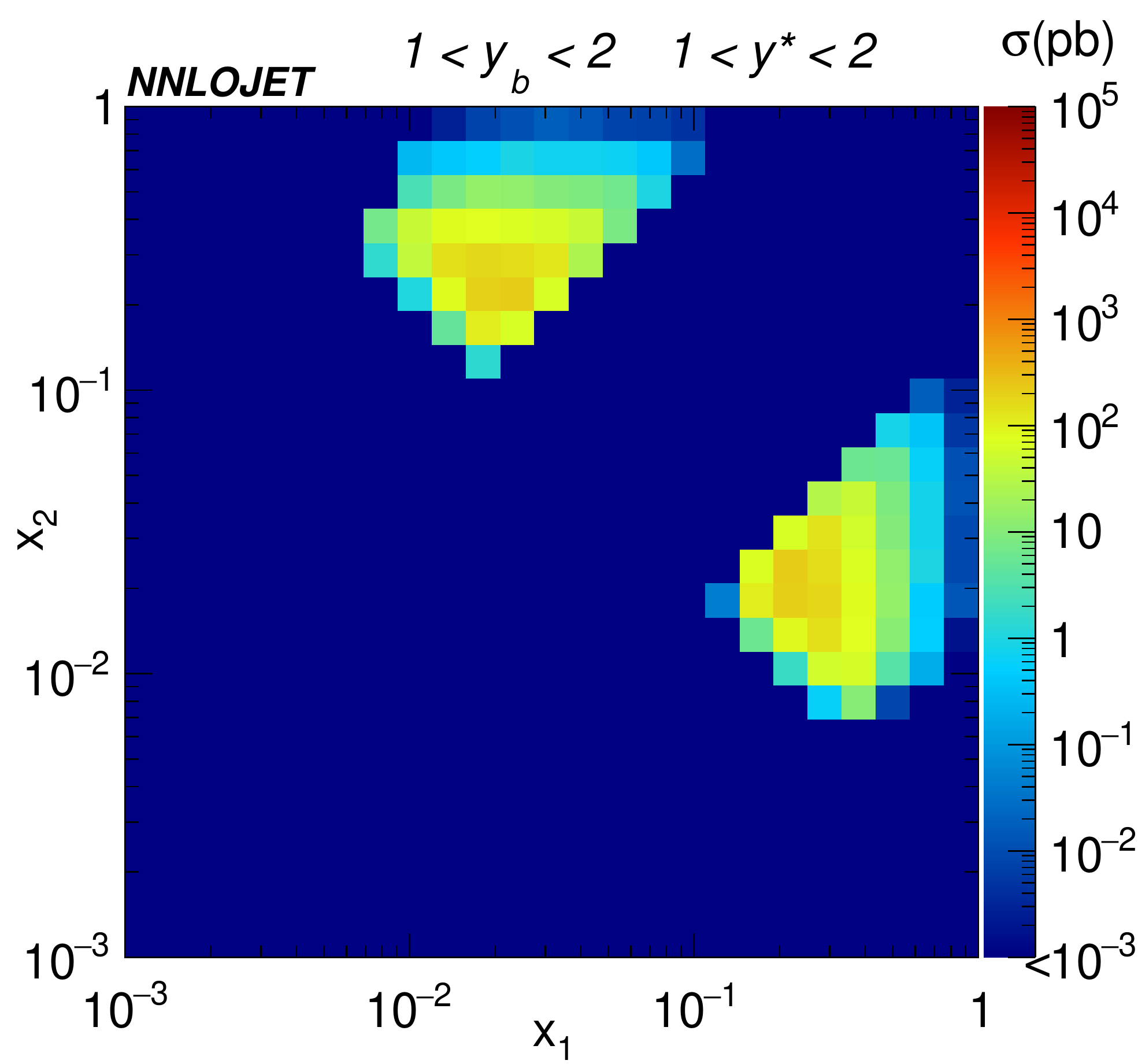}\hspace{0.5cm}
    \includegraphics[width=0.28\textwidth]{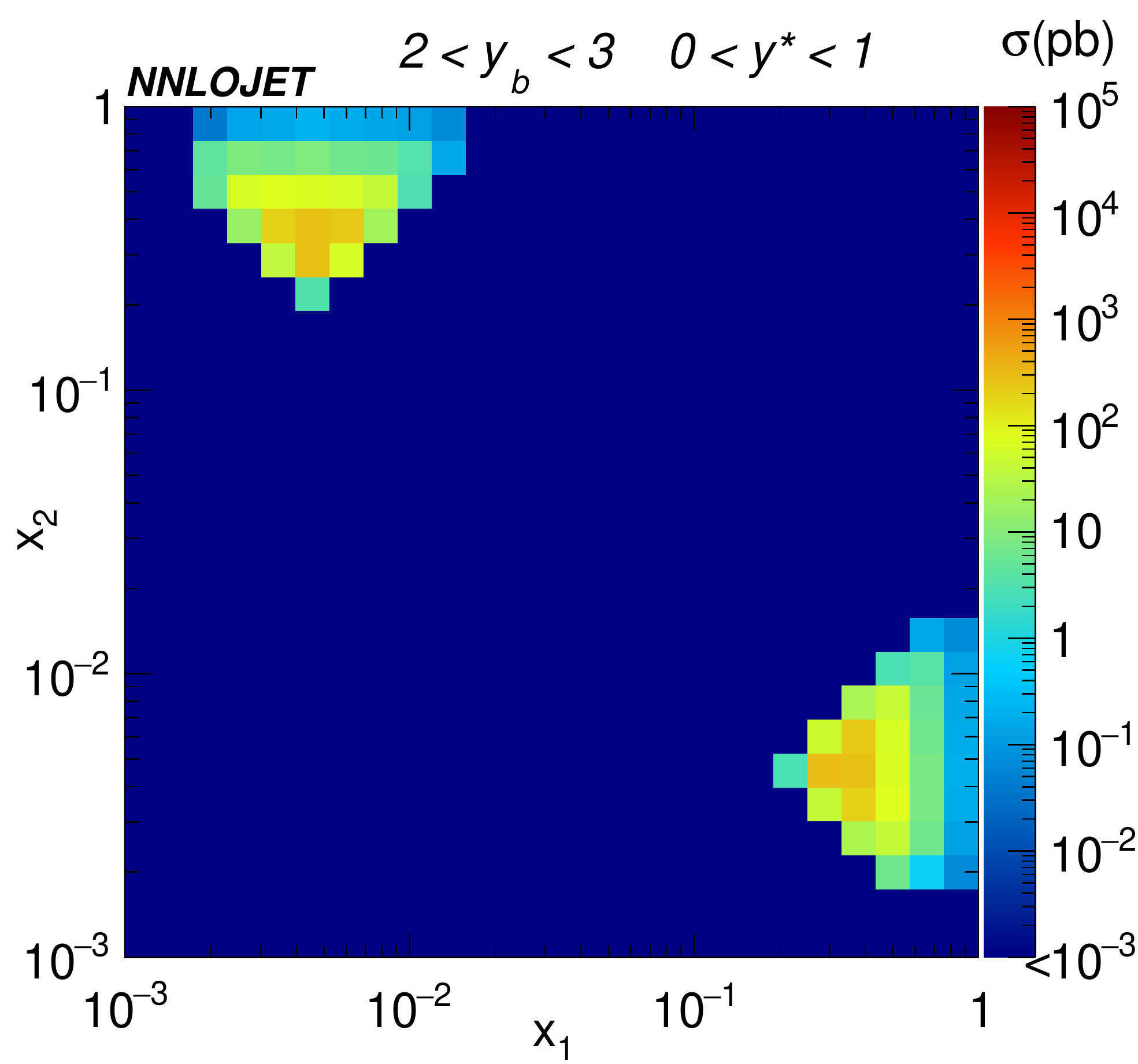}  
  \caption{Allowed kinematical regions at LO in the triple differential dijet inclusive cross section (in pb) at $\sqrt{s}=8$~TeV in the partonic fraction $x_{1},x_{2}$ plane for the jet $p_{T}$ cuts of the CMS measurement.
  }
  \label{fig:x1x2kin}
\end{figure*}

The CMS 8~TeV measurement~\cite{Sirunyan:2017skj} is performed in six different regions in $(y^{*},y_b)$, each corresponding to 
different types of event topologies and probing different aspects of the partonic structure of the colliding protons. 
In Fig.~\ref{fig:x1x2kin}, a density plot in the ($x_{1},x_{2}$) plane of the triple-differential cross section
for the six event topologies considered in the CMS study is shown.
The CMS analysis~\cite{Sirunyan:2017skj} also performs 
a detailed study of the constraints on PDFs that can be derived from the measurement data. These turn out to be inherently limited by the precision of the 
theoretical description of the underlying hard scattering processes available. 

The theoretical predictions for
the jet cross section are obtained in perturbative QCD, as a convolution of the parton distribution functions for the incoming particles and the parton--parton hard scattering cross section. 
The previous state of the art, as used in Ref.~\cite{Sirunyan:2017skj}, were predictions including next-to-leading order (NLO)  QCD~\cite{Giele:1994gf,Ellis:1992en,Nagy:2001fj,Alioli:2010xa} and 
electroweak (EWK) corrections~\cite{Dittmaier:2012kx,Campbell:2016dks,Frederix:2016ost} for the dijet production cross section. 
At this level of accuracy, scale uncertainties and missing higher order corrections in the theoretical calculation significantly limit the achievable precision in the determination of the shape and normalization of the triple differential cross section. 
To improve the perturbative QCD description of this process, we present in this Letter for the first time
a computation of the NNLO corrections to the triple-differential dijet cross section at the LHC.

Our calculation is performed in the \NNLOJET framework, employing the antenna subtraction method~\cite{GehrmannDeRidder:2005cm,Daleo:2006xa,Currie:2013vh} 
to remove all unphysical infrared singularities from the matrix
elements, which we take at leading color in all partonic subprocesses at NNLO, while keeping the full color dependence at lower orders. The same setup was used for the 
calculation of the NNLO corrections to inclusive jet~\cite{Currie:2016bfm,Currie:2018xkj} and dijet production~\cite{Currie:2017eqf}. 
We use the MMHT2014 NNLO parton distribution functions~\cite{Harland-Lang:2014zoa} with $\alpha_s(M_Z)$=0.118 for all predictions at LO, NLO, and NNLO to emphasize the role
of the perturbative corrections at each successive order.

The combined nonperturbative (NP) contributions from hadronization and the underlying event, modeled through multiple parton interactions, 
are not included in the predictions at the parton-level, but have been derived from parton shower predictions at NLO in Ref.~\cite{Sirunyan:2017skj}. The corresponding NP effects have been found
to be at most 10\% for the lowest $p_{T,\textrm{avg}}$ bins and negligible above 1~TeV. A recent study~\cite{Bellm:2019yyh} has shown that for the $R=0.7$ cone
size, there is an excellent agreement for the parton-level cross section between fixed-order and NLO-matched results. For this reason, 
we will take into acount the NP effects obtained in Ref.~\cite{Sirunyan:2017skj} as a multiplicative factor in each bin of the parton-level NNLO cross section,
labelling the results as NNLO$\otimes$NP. 

The contribution from EWK effects from virtual exchanges of massive $W$ and $Z$ bosons have been computed in Ref.~\cite{Dittmaier:2012kx}. These are smaller than 3\%
below 1 TeV and reach 8\% for the highest $p_{T,\textrm{avg}}$.  Using the results from Ref.~\cite{Dittmaier:2012kx}, EWK corrections
are applied multiplicatively to the QCD calculation for the central scale choice and we label the corresponding prediction NNLO$\otimes$NP$\otimes$EWK.

At any given fixed order in perturbation theory, the predictions retain some dependence on the unphysical renormalization and factorization scales.
An assessment of the scale uncertainty of the calculation at NLO and NNLO is obtained from independent variations of the renormalization or factorization scales by a factor of 2 around
an arbitrary central scale choice. For the process at hand, the production of dijet events at high-$p_{T}$ have as a natural scale (see Ref.~\cite{Currie:2017eqf} for a detailed study), 
the mass of the dijet system $m_{jj}=(p_{j1}+p_{j2})^2$,
which closely approximates the four-momentum transferred in the interaction.

\begin{figure}[b]
\centering
\includegraphics[width=0.5\textwidth]{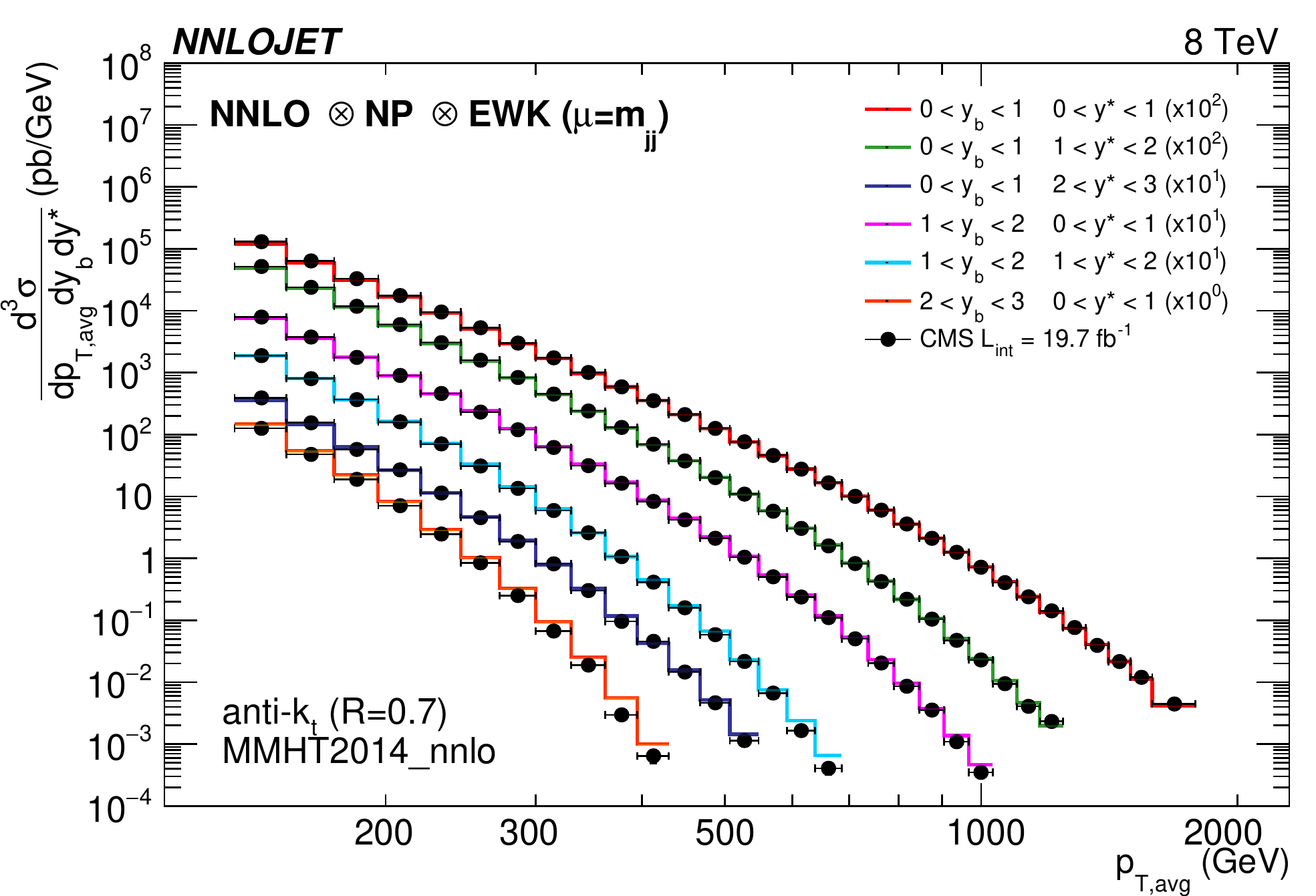}
\caption{The triple differential dijet cross section as a function of $p_{T,\textrm{avg}}$ for the six bins of $y^{*}, y_{b}$ at NNLO with central scale choice $\mu=m_{jj}$, compared to CMS
8~TeV 19.7~fb$^{-1}$ data.}
\label{fig:xsec}
\end{figure}

In Fig.~\ref{fig:xsec}, we present the absolute dijet triple-differential cross section at NNLO$\otimes$NP$\otimes$EWK as a function of $p_{T,\textrm{avg}}$, $y^{*}$ and $y_{b}$, which we compare to LHC data. 
The fiducial cuts of the measurement include all events with at least two jets with absolute rapidity up to $|y|\le 5.0$ recorded 
in the CMS 8~TeV 19.7~fb$^{-1}$ data, where the two jets leading in $p_{T}$ are required to have $p_{T}\ge 50$~GeV and $|y|\le 3.0$. 
The jet reconstruction in both experimental analysis and theory prediction uses the anti-$k_{T}$ jet algorithm~\cite{Cacciari:2008gp}
with radius parameter R=0.7. We observe an excellent agreement in the description of the data by the NNLO prediction over 7 orders of magnitude.

\begin{figure}[b]
\centering
\includegraphics[width=0.45\textwidth]{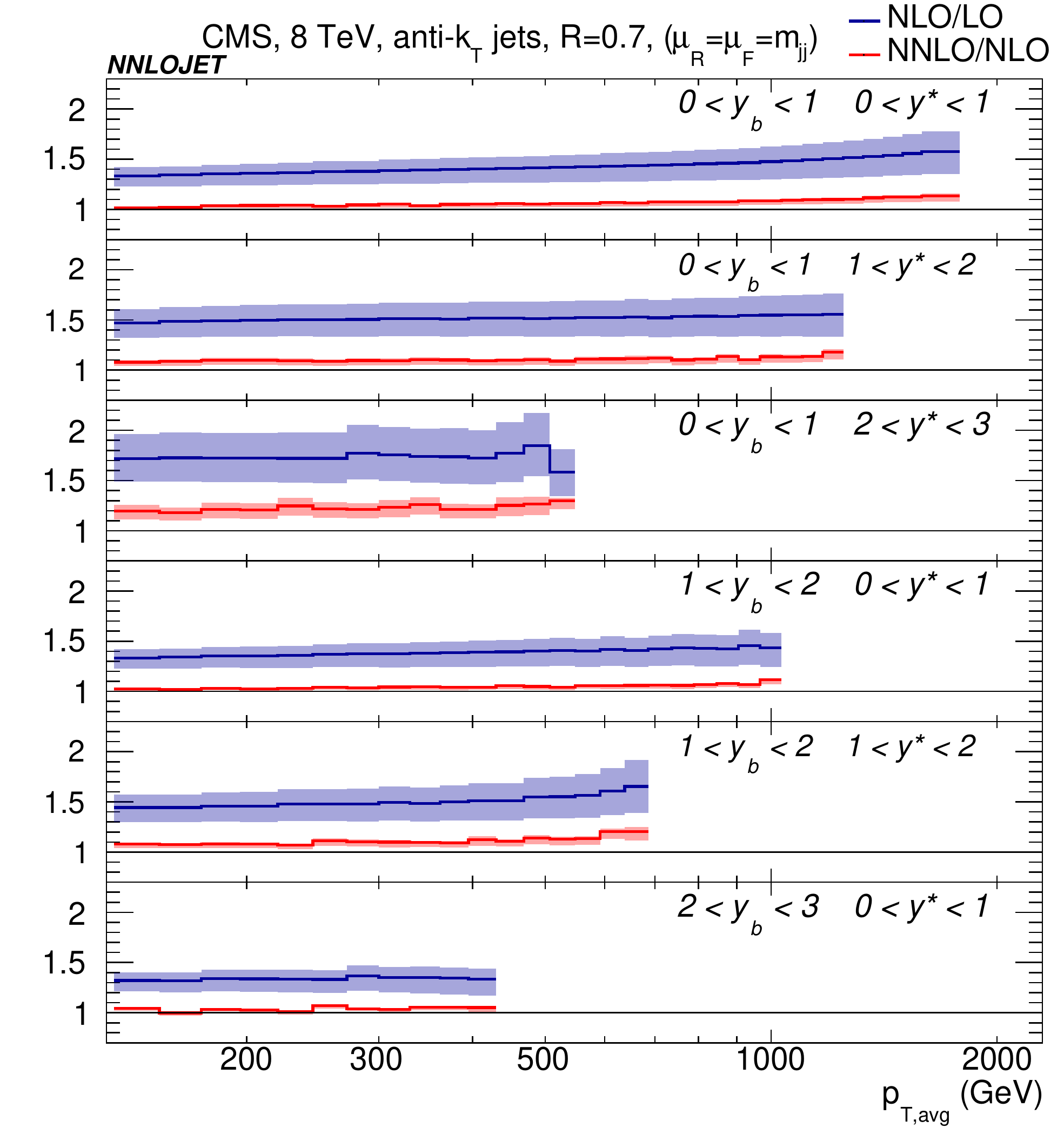}
\caption{NLO/LO (blue) and NNLO/NLO (red) $K$ factors triple differential in $p_{T,\textrm{avg}}$, $y^{*}$ and $y_{b}$. Bands represent
the scale variation of the numerator. NNLO PDFs are used for all predictions.}
\label{fig:kfac}
\end{figure}

An assessment of the impact of the newly computed NNLO contribution can be seen in Fig.~\ref{fig:kfac}, where we show explicitly the ratio between the NNLO prediction and the NLO result (in red), together
with the ratio between the NLO cross section and the LO result (in blue). The size of the NNLO corrections varies significantly as a function of  $p_{T,\textrm{avg.}}$ and the applied cuts on $y^{*}$ and $y_{b}$.

For the central $y_{b}$ slice, and at small $y^{*}$, we observe NNLO effects that range between a few percent at low $p_{T,\textrm{avg.}}$, rising to 15\% at large $p_{T,\textrm{avg}}$ (first panel). For larger
$y^{*}$ slices, i.e., smaller scattering angles, the NNLO effects increase reaching up to 20\% across the entire $p_{T,\textrm{avg}}$ range in the $2 < y^* < 3$ rapidity slice. On the other hand, for
the longitudinally boosted topologies, i.e., at larger values of $y_{b}$, the NNLO effects are smaller (as can be seen in the lower three panels in Fig.~\ref{fig:kfac}). 
They range between a few percent at low $p_{T,\textrm{avg}}$ and increase slightly to 10\%, for  the smaller scattering angle topologies, typically found at large  $y^*$.

We also observe a good convergence of the perturbative expansion with NNLO effects smaller in magnitude with respect to the NLO correction at the previous order, 
demonstrating that the variables chosen to describe the measurement are stable and infrared safe, and allow us to reliably predict the triply differential cross section over the whole kinematical range
in perturbation theory. In this respect, we note that the NNLO correction changes both the shape and normalization of the NLO result, and can therefore have a significant impact towards obtaining 
better constraints on parton distribution functions using data from the triple differential dijet measurement. 

 \begin{figure*}[t]
  \centering
    \includegraphics[width=0.32\textwidth,page=1]{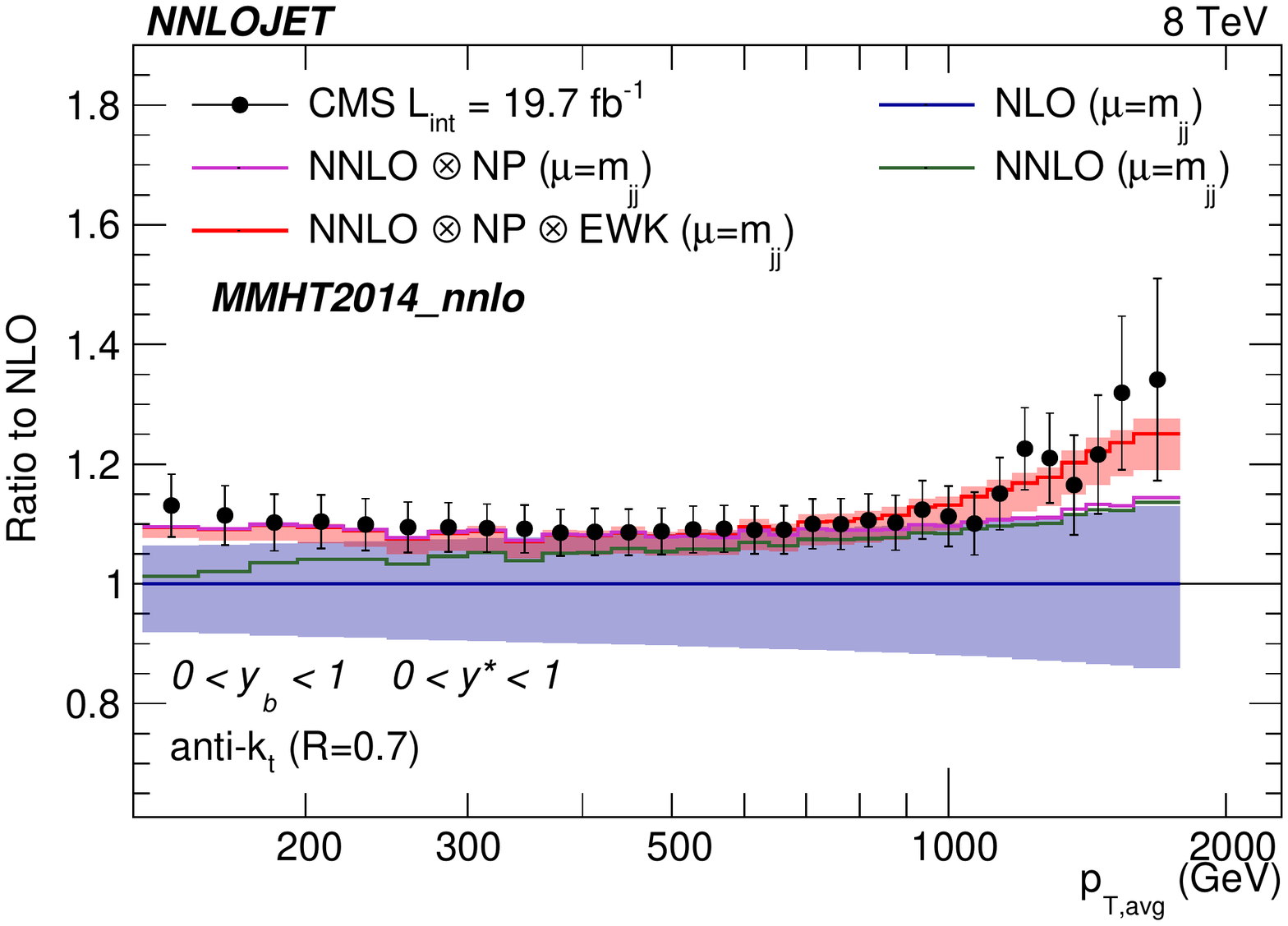}
    \includegraphics[width=0.32\textwidth,page=2]{Fig5R07mu-mjj_MMHT2014nnlo}
    \includegraphics[width=0.32\textwidth,page=3]{Fig5R07mu-mjj_MMHT2014nnlo}
    \includegraphics[width=0.32\textwidth,page=4]{Fig5R07mu-mjj_MMHT2014nnlo}
    \includegraphics[width=0.32\textwidth,page=5]{Fig5R07mu-mjj_MMHT2014nnlo}
    \includegraphics[width=0.32\textwidth,page=6]{Fig5R07mu-mjj_MMHT2014nnlo}    
  \caption{The NLO (blue) and NNLO (green) theory predictions and CMS data normalized to the NLO central value. Parton-level predictions corrected for
  nonperturbative (NP) effects and combined NP and Electroweak effects (EWK), implemented as a multiplicative factor to the NNLO result, are shown shown in pink and red.
  The shaded bands shown for the NLO and the NNLO$\otimes$NP$\otimes$EWK predictions represent the variation of the theoretical scales in the numerator by factors of 0.5 and 2.
  The uncertainty in the data is the total experimental uncertainty, including systematic and statistical uncertainties added in quadrature.
  }
  \label{fig:MMHTnnlo}
\end{figure*}

Owing to the large dynamical range covered by the CMS measurement, a quantitative comparison between data and theory  is best performed by taking ratios, and we 
use the  NLO QCD parton-level prediction as reference value. The resulting ratios  to the CMS data~\cite{Sirunyan:2017skj} and to the NNLO predictions at parton-level, and 
with NP and EWK corrections applied are displayed in Fig.~\ref{fig:MMHTnnlo}.
The results show
a significant improvement in the description of the data at NNLO with respect to NLO, for the entire kinematical range of the measurement, 
in particular for the central $y_{b}$ slice (top three panels).  

In the lower three panels of  Fig.~\ref{fig:MMHTnnlo}, which correspond to the highly boosted and PDF-sensitive kinematical bins, 
we observe that the data tends to be below the central value of the MMHT2014 PDF set even at NNLO, in particular at large values of $p_{T,\textrm{avg.}}$. In this region, 
which is sensitive to the scattering of large-$x$ parton on a low-$x$ parton (see Fig.~\ref{fig:x1x2kin}), the PDFs suffer from large uncertainties. Our results at NNLO suggest 
that the understanding of the high-$x$ behaviour of the PDFs can be improved upon by including measurements of  triple differential dijet distributions in 
future global PDF determinations. 

In this Letter, we computed the second-order QCD corrections to the triple-differential dijet production cross section at hadron colliders. 
Our results substantially improve  the theoretical description of this benchmark observable, with theory uncertainties now being comparable or 
lower than experimental errors, while better explaining kinematical shapes. Our newly derived results will enable the usage of dijet measurements in 
precision studies of the  partonic structure of the colliding hadrons. 

 The authors thank Xuan Chen, James Currie,
  Marius H\"ofer, Imre Majer, Jan Niehues, Duncan Walker and James Whitehead for useful discussions and their many contributions to the \NNLOJET code. 
This research was supported in part by the UK Science and Technology Facilities Council under Contract No. ST/G000905/1, by the Swiss National Science Foundation (SNF) under Contracts No. 200021-172478 and No. 200020-175595, by the Research Executive Agency (REA) of the European Union through the ERC Consolidator Grant HICCUP (614577) and the ERC Advanced Grant MC@NNLO (340983), by the Funda\c{c}\~{a}o para a Ci\^{e}ncia e Tecnologia (FCT-Portugal), under Projects No. UID/FIS/00777/2019, No. CERN/FIS-PAR/0022/2017, and COST Action CA16201 PARTICLEFACE. J.P. gratefully acknowledges the hospitality and financial support of the CERN theory group where work on this paper was conducted.
     
\bibliography{ref3d}
     
\end{document}